\documentclass[twocolumn,nolinenumbers,trackchanges]{aastex631}
\graphicspath{{./}{figures/}}
\usepackage{amsmath}
\usepackage{hyperref}
\usepackage{xcolor}
\usepackage{graphicx}
\begin{document}

\title{Determining the Origin of Very-high-energy Gamma Rays from Galactic Sources by Future Neutrino Observations}

\author{Bo-Heng Song}
\affiliation{Department of Astronomy, School of Physics, Huazhong University of Science and Technology, Wuhan 430074, China;}

\author[0000-0001-8510-2513]{Tian-Qi Huang}
\affiliation{Key Laboratory of Particle Astrophysics and Experimental Physics Division and Computing Center, Institute of High Energy Physics, Chinese Academy of Sciences,
100049 Beĳing, China;}
\affiliation{Tianfu Cosmic Ray Research Center, 610000 Chengdu, Sichuan, China}

\author[0000-0003-4976-4098]{Kai Wang}
\affiliation{Department of Astronomy, School of Physics, Huazhong University of Science and Technology, Wuhan 430074, China;}

\correspondingauthor{Kai Wang}
\email{kaiwang@hust.edu.cn}

\begin{abstract}

Recently, the Large High Altitude Air Shower Observatory (LHAASO) identified 12 $\gamma$-ray sources emitting gamma rays with energies above 100 TeV, making them potential PeV cosmic-ray accelerators (PeVatrons). Neutrino observations are crucial in determining whether the gamma-ray radiation process is of hadronic or leptonic origin. In this paper, we study three detected sources, LHAASO J1908+0621, LHAASO J2018+3651, and LHAASO J2032+4102, which are also the most promising galactic high-energy neutrino candidate sources with the lowest pre-trial p-value based on the stacking searches testing for excess neutrino emission by IceCube Neutrino Observatory. We study the lepto-hadronic scenario for the observed multiband spectra of these LHAASO sources considering the possible counterpart source of the LHAASO sources. The very-high-energy gamma rays are entirely attributed to the hadronic contribution, therefore the most optimistic neutrino flux can be derived. Then, we evaluate the statistical significance (p-value) as a function of observation time of IceCube and the next-generation IceCube-Gen2 neutrino observatory respectively. Our results tend to disfavor that all gamma rays above $100\,\rm GeV$ from LHAASO J1908+0621 are of purely hadronic origin based on current IceCube observations, but the purely hadronic origin of gamma rays above $100\,\rm TeV$ is still possible. By IceCube-Gen2, the origin of gamma rays above $100\,\rm TeV$ from LHAASO J1908+0621 can be further determined at a $5\sigma$ significance level within a running time of $\sim 3$ years. For LHAASO J2018+3651 and LHAASO J2032+4102, the required running time of IceCube-Gen2 is $\sim 10$ years ($3\sigma$) and $\sim 10$ years ($5\sigma$), respectively. Future observations by the next-generation neutrino telescope will be crucial to understanding the particle acceleration and radiation processes inside the sources.

\end{abstract}

\keywords{High-energy astrophysics; Gamma-rays; Supernova remnants; Pulsars; Neutrino}

\section{Introduction} \label{sec:intro}

The origin of high-energy cosmic rays (CRs) has been a long-standing question in particle astrophysics.
%Cosmic rays are primarily composed of 90 $\%$ protons, about 8-9 $\%$ helium nuclei, and trace amounts of heavier elements, and they travel through space at relativistic speeds. 
The cosmic ray spectrum is typically described by a power law with an index of $\sim 2.7$ up to the so-called ``knee" at around 3 PeV, beyond which the spectrum softens \citep{abbasi2018cosmic}. CR composition changes from light (protons and helium) to heavier as a function of energy near the knee and is known to be dominated by protons~\citep{2013FrPhy...8..748G}. This suggests the existence of powerful astrophysical proton accelerators in our galaxy, which can accelerate protons to energies up to a PeV, commonly referred to as "PeVatrons". The potential galactic PeVatrons can be identified by the very-high-energy (VHE, $> 100\,\rm GeV$) gamma-ray detection and have been explored by ground-based telescopes, such as \href{https://www.mpi-hd.mpg.de/hfm/HESS/}{H.E.S.S.} (High Energy Stereoscopic System)~\citep{hess2016, abdalla2018characterising}, \href{https://magic.mpp.mpg.de/}{MAGIC} (Major Atmospheric Gamma Imaging Cherenkov)~\citep{acciari2020magic}, \href{https://www.hawc-observatory.org/}{HAWC} (High-Altitude Water Cherenkov)~\citep{Albert_2020}, \href{https://www.hawc-observatory.org/} and \href{http://english.ihep.cas.cn/lhaaso/}{LHAASO} (Large High Altitude Air Shower Observatory)~\citep{cao2021ultrahigh,doi:10.1126/science.abg5137,cao2023first}. Especially, LHAASO can capture gamma rays with energies from hundreds of GeV to exceeding PeV \citep{lhaaso2010future}, the sources of which have strong possibilities of being galactic PeVatrons. However, the origin of VHE gamma-rays is still in debate. The VHE gamma rays can be produced through the decay of pions which are generated by hadronic processes between the accelerated high-energy cosmic rays and the surrounding medium. An alternative scenario is leptonic processes, such as inverse Compton scattering and bremsstrahlung of the accelerated high-energy electrons, which can also produce high-energy gamma rays. Therefore, confirming the gamma-ray origin is crucial to identifying the composition of accelerated particles and the radiation processes for PeVatrons.

%A significant probe to determine the VHE gamma-ray origin is the high-energy neutrino, 
Neutrinos are a significant probe in determining the origin of VHE gamma rays, which are concomitantly produced with gamma rays of hadronic origin. Therefore, the detection or non-detection of high-energy neutrinos can be a diagnosis of the hadronic or leptonic origin of VHE gamma rays. Very recently, neutrino emission from the galactic plane has been identified at the $4.5\sigma$ level of significance by IceCube Neutrino Observatory~\citep{doi:10.1126/science.adc9818}, which implies that galactic sources can generate high-energy neutrinos. In addition, for the LHAASO sources, \cite{2023ApJ...945L...8A} conducted stacking searches testing for excess neutrino emission from 12 LHAASO sources which are identified with emissions above 100 TeV~\citep{cao2021ultrahigh} and thought to be PeVatron candidates. Although no significant neutrino emissions were found, three LHAASO sources, i.e., LHAASO J1908+0621, LHAASO J2018+3651, and LHAASO J2032+4102, present the lowest p-values, making them as the promising sources to identify the possible neutrino emission in the future and then judge the origin of VHE gamma rays.

Due to the complex spatial morphology of three LHAASO sources, the gamma-ray counterparts for these sources are still uncertain, which can be supernova remnants (SNRs), pulsar wind nebulae (PWNe), or young massive star clusters (YMCs). SNRs are widely thought of as the primary galactic cosmic-ray sources, and the particles can be accelerated by diffusive shock acceleration in their forward shocks generated by the interaction of supernova ejecta with the interstellar medium (ISM). The production of gamma rays and high-energy neutrinos from interactions of accelerated CR protons and nuclei with ambient medium \citep{gabici2007searching}. CR protons can also be accelerated and trapped in PWNe and YMCs and then produce gamma rays and neutrinos \citep{di2017revised}.

In this paper, we collect the multiband spectra observed from the direction of these sources and consider the possible counterpart sources of the LHAASO sources. With the proposed theoretical scenario, the multiband spectral modeling is implemented and VHE gamma rays are mainly attributed to the hadronic process. With the most optimistic neutrino production in the sources, we evaluate the statistical significance as a function of observation time for three sources, i.e., LHAASO J1908+0621, LHAASO J2018+3651, and LHAASO J2032+4102, using the IceCube and next-generation IceCube-Gen2 neutrino observatory respectively. 

The remaining part of this paper is organized as follows. In section \ref{sec:MODEL}, we obtain the SED of sources through the lepto-hadronic scenario. In section \ref{sec:neu}, we calculate the corresponding neutrino SED from the hadronic interaction and compare the calculation with the sensitivity of the IceCube-Gen2 observatory. In section \ref{sec:future}, we estimate the statistical significance of neutrino signals from LHAASO sources using both the IceCube and the proposed IceCube-Gen2. Finally, section \ref{sec:discussion and summary} is the discussion and summary.

\section{THEORETICAL MODELING} \label{sec:MODEL}
\subsection{LHAASO J1908+0621}

Although the nature of LHAASO J1908+0621 is still unknown, it is considered one of the most promising PeVatron candidates in the galaxy due to its extended bright TeV emissions~\citep{cao2021ultrahigh}. LHAASO J1908+0621 is spatially associated with a middle-aged and shell-like supernova remnant SNR G40.5-0.5~\citep[20-40 kyr;][]{downes1980g40}, an energetic gamma-ray pulsar PSR J1907+0602 (age of 19.5 kyr, distance of $3.2 \pm 0.6\,\rm kpc$, spin-down luminosity of $\sim 2.8 \times 10^{36}\,\rm erg/s$) \citep{2010ApJ...711...64A,li2021investigating}, and an energetic radio pulsar PSR J1907+0631 (age of 11 kyr, distance of 7.9 kpc, spin-down luminosity of $\sim 5 \times 10^{35}\,\rm erg/s$) \citep{lyne2017timing}. The distance estimates of SNR G40.5-0.5 place it at a distance of $3.5\,\rm kpc$ using CO observations \citep{yang2006molecular} or a more distant position of 5.5-8.5 kpc~\citep{downes1980g40} or 6.1 kpc~\citep{case1998new} using the $\Sigma$-D relation. Besides, an unidentified GeV source, 4FGL J1906.2+0631, is also spatially associated with LHAASO J1908+0621 as reported in \cite{li2021investigating}. In addition, by analyzing the distribution of the CO gas, the molecular clouds (MCs) spatially correlated with SNR G40.5-0.5 and the gamma-ray emission are identified~\citep{duvidovich2020radio,li2021investigating}.

The origin of the gamma-ray emission from LHAASO J1908+0621 is still under debate due to its complex spatial morphology. 
%In principle, the pulsar PSR J1907+0631 is unlikely responsible for the gamma-ray emission due to its location lacking gamma-ray emission and its significant offset from the position of the gamma-ray emission~\citep{duvidovich2020radio}.
In principle, PSR J1907+0631 can potentially power the entire TeV emissions, however, considering that it is not a gamma-ray pulsar, we neglect its contributions to gamma-ray emissions~\citep{duvidovich2020radio,lyne2017timing}.
Other possible scenarios can be concluded as follows. A leptonic component from the PWN powered by PSR J1907+0602 was initially proposed as the origin of VHE gamma rays~\citep{2010ApJ...711...64A}. The combination of leptonic and hadronic scenarios has been proposed as well for the origin of gamma-ray emission in this region~\citep{duvidovich2020radio,2021MNRAS.505.2309C,li2021investigating,2022ApJ...928..116A,de2022exploring}. The hadronic component is usually related to SNR G40.5-0.5. As suggested in \cite{de2022exploring}, we consider the $pp$ interaction of escaped protons accelerated by the shock of SNR G40.5-0.5 with the materials of MCs in the MC region, and a leptonic component from SNR G40.5-0.5, which is located at a distance of 2.37 kpc, similar to \cite{albert2021spectrum} and \cite{cao2023first}.

\begin{figure*}[ht!]
\gridline{\fig{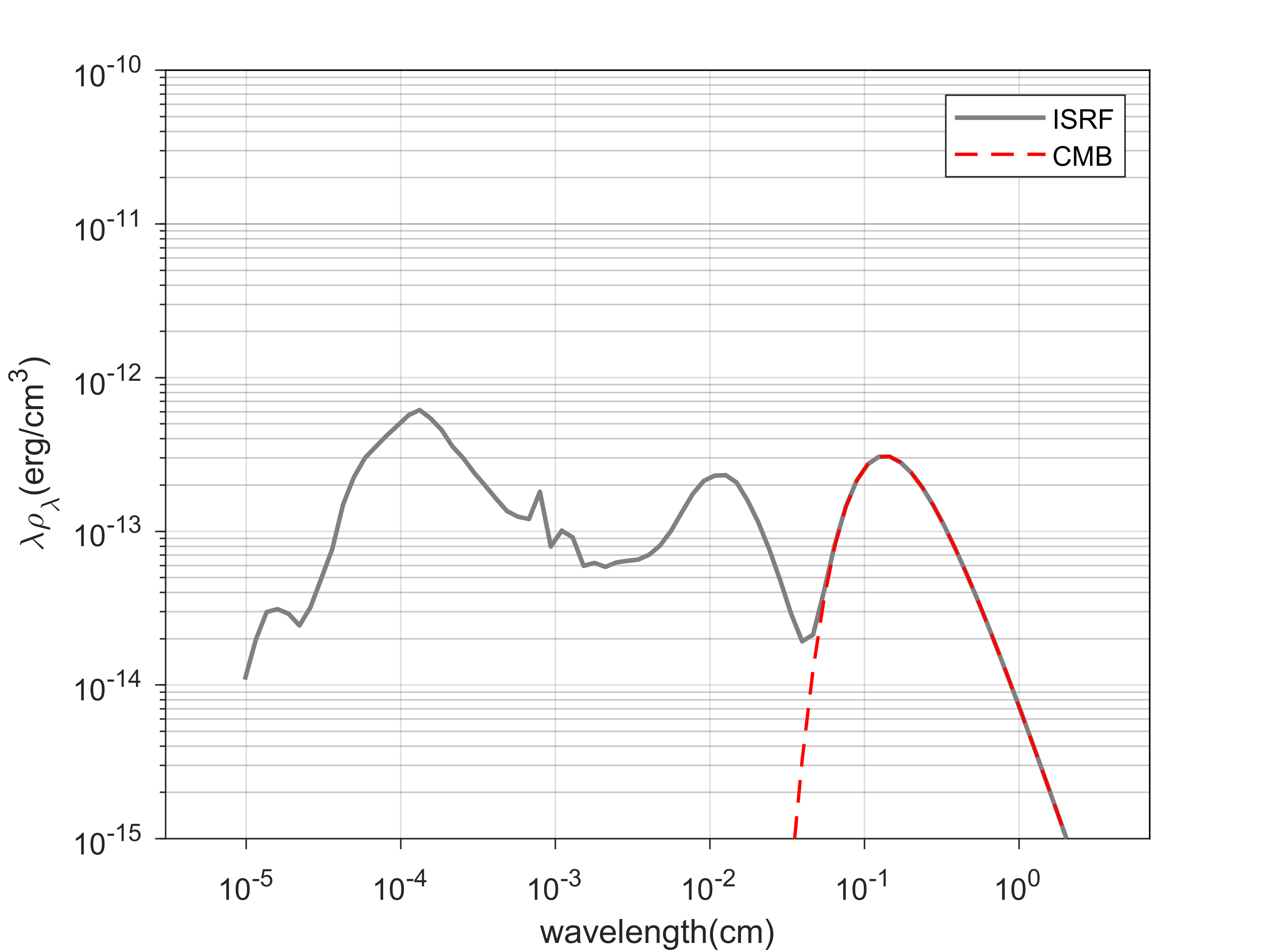}{0.3\textwidth}{LHAASO J1908+0621}
          \fig{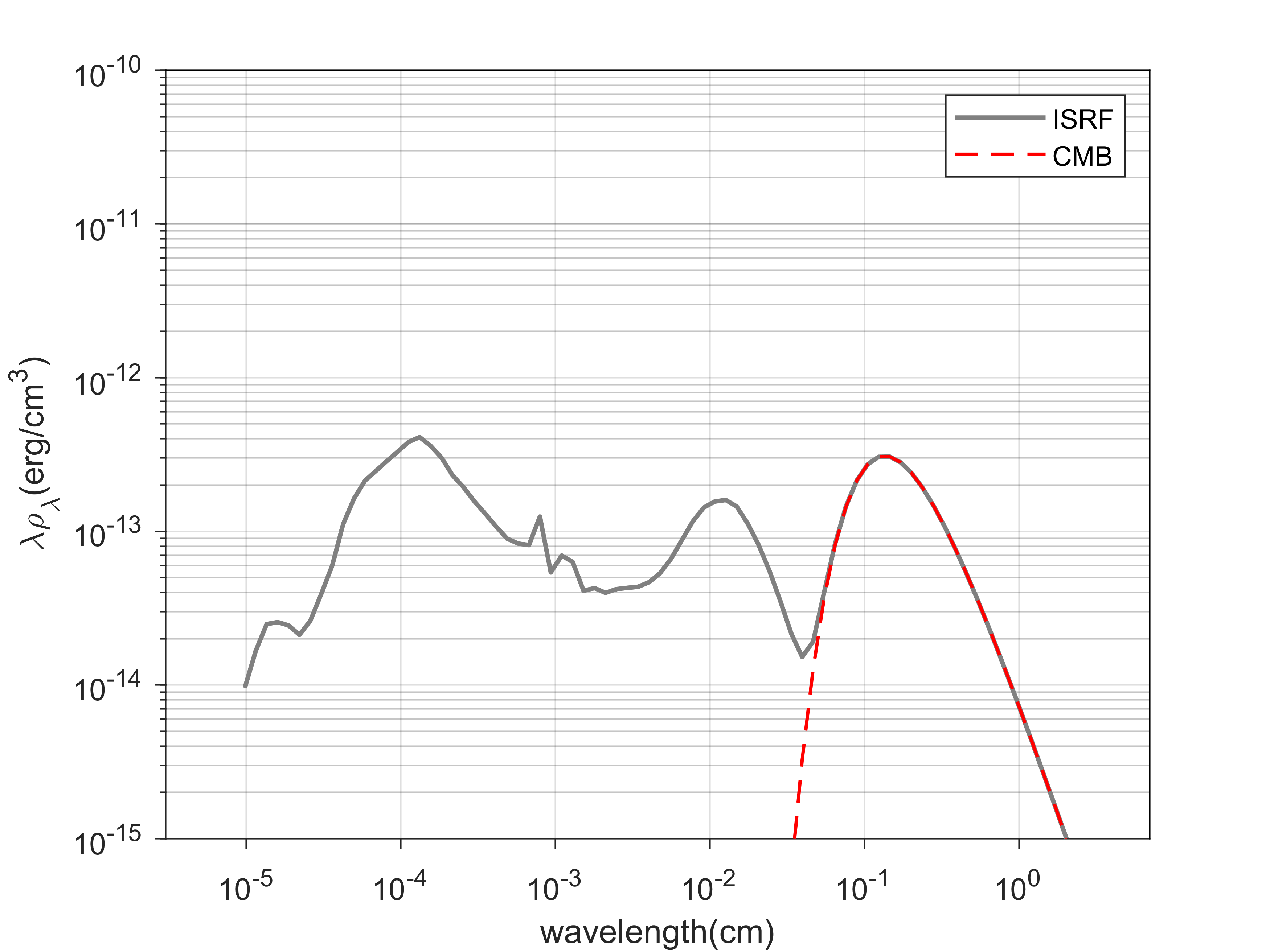}{0.3\textwidth}{LHAASO J2018+3651}
          \fig{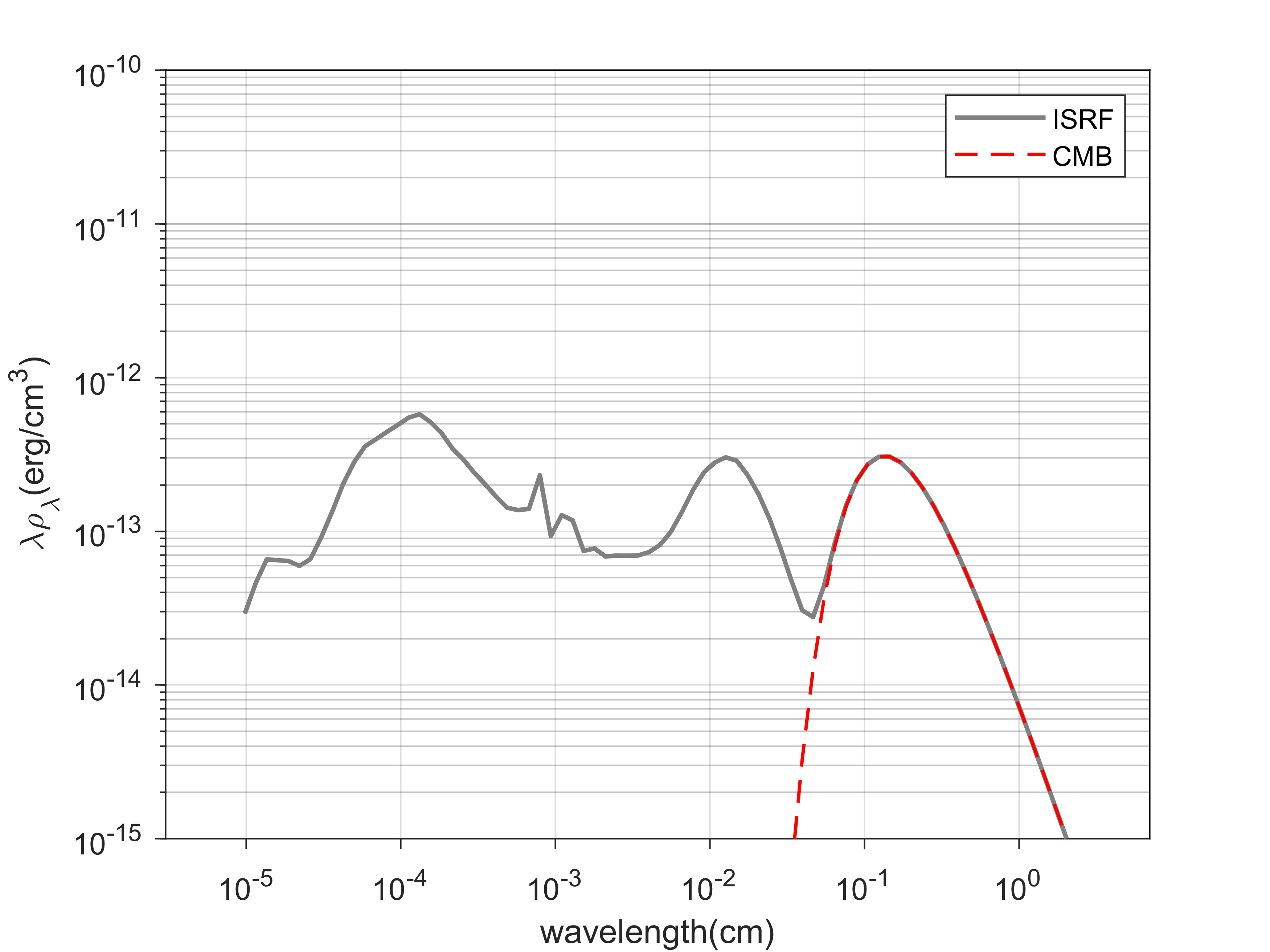}{0.3\textwidth}{LHAASO J2032+4102}
          }
          \caption{Radiation field of each source. The red dashed line is the cosmic microwave background (CMB) energy density. The grey curve is the sum of CMB and interstellar radiation field (ISRF) at diverse galactocentric positions.
\label{fig1}}
\end{figure*}   

\begin{figure*}[ht!]
\gridline{\fig{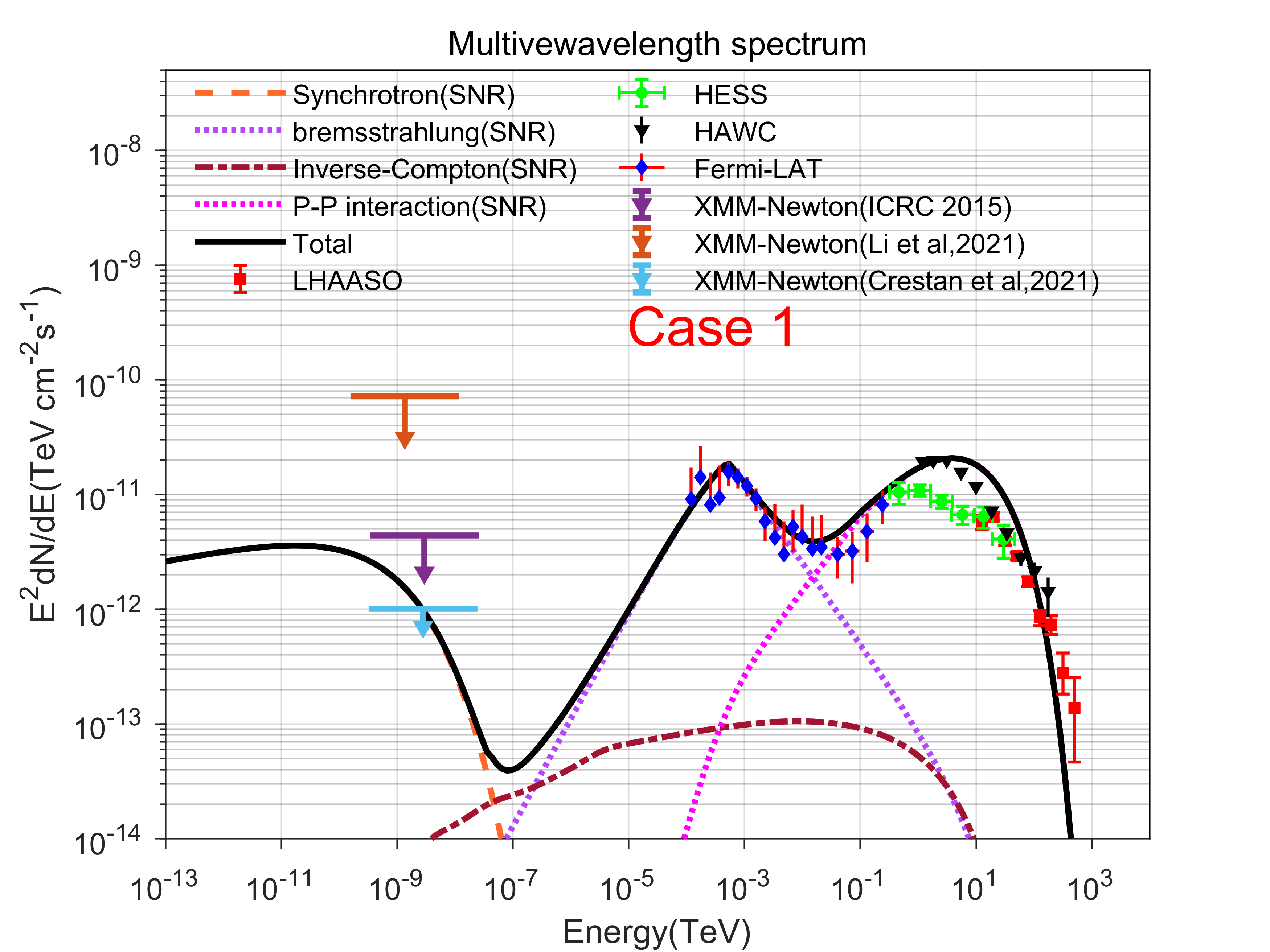}{0.49\textwidth}{(a)}
          \fig{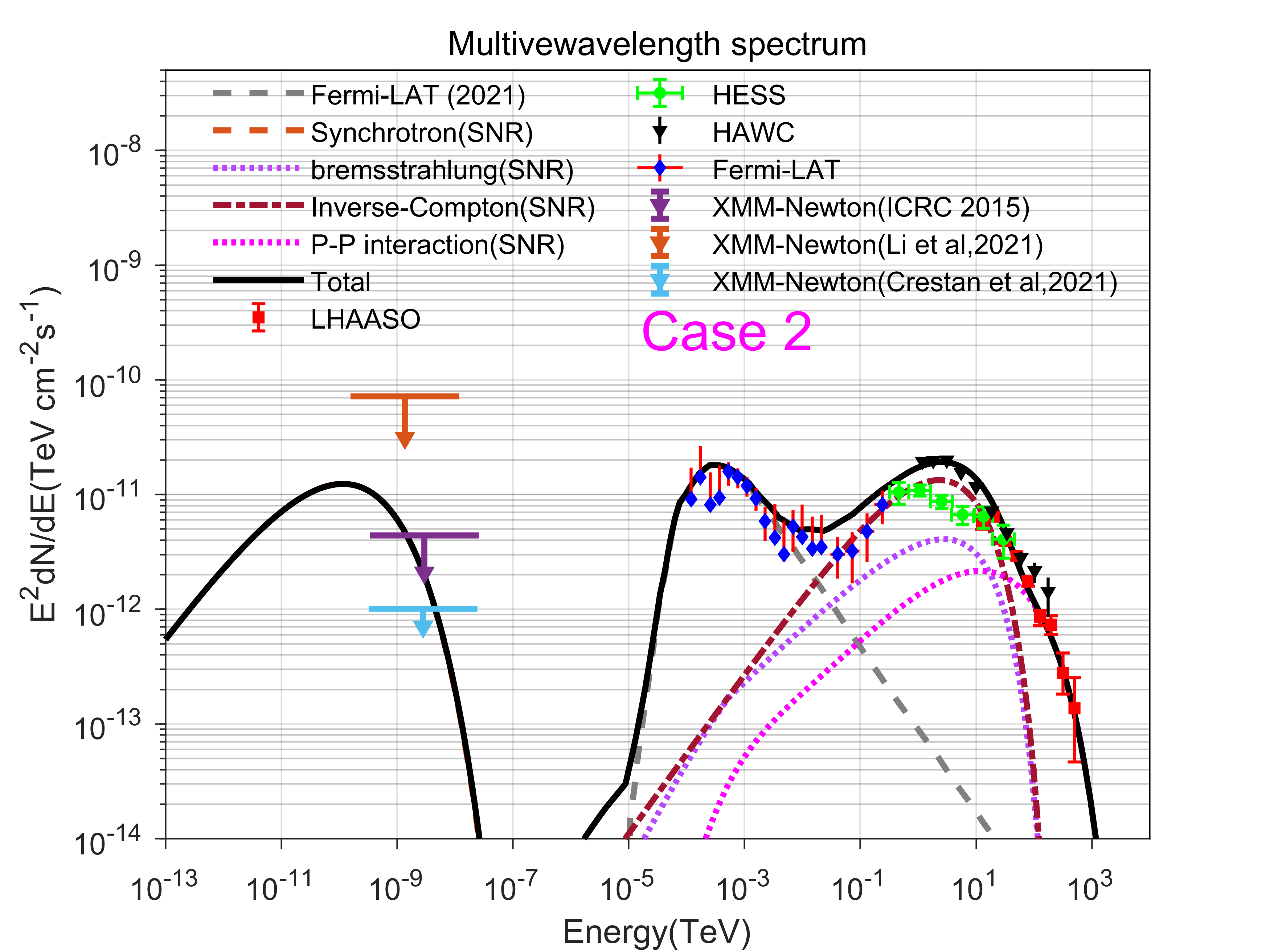}{0.49\textwidth}{(b)}
          }
          \caption{Multi-wavelength SED and fitting for LHAASO J1908+0621. Datapoints obtained from different observations by LHAASO (red) \citep{cao2021ultrahigh}, H.E.S.S. (green) \citep{aharonian2009detection}, HAWC (black) \citep{abeysekara2020multiple}, \textit{Fermi}-LAT (blue \citep{li2021investigating} are shown in the figure. The XMM-Newton upper limit obtained from \cite{pandel2015investigating} is shown in purple. The XMM-Newton upper limits obtained from \cite{li2021investigating} and \cite{crestan2021multiwavelength} are shown in brownness and light blue respectively. (a) The synchrotron (Orange dashed), bremsstrahlung (purple dotted), and IC (brown dot-dashed) components are shown. Also the pink dotted line corresponds to the hadronic component from SNR G40.5-0.5 shown. The sum of these components is shown with a black solid line. (b) The gray dashed line shows an additional component of the MeV-GeV range \citep{li2021investigating}. The synchrotron (Orange dashed, covered by the solid black line), bremsstrahlung (purple dotted), and IC (brown dot-dashed) components are shown. Also, the pink dotted line corresponds to the hadronic component from SNR G40.5-0.5 shown. The sum of these components is shown with a black solid line as well. The model parameters of the two panels are summarized in Table \ref{tab1}.
\label{fig2}}
\end{figure*}

The acceleration mechanisms of cosmic rays have been studied for a long time. One of the most popular acceleration mechanisms is the diffusive shock acceleration related to the supernova remnants (\cite{fermi1949origin, drury1983introduction, schure2012diffusive}). Charged particles would be accelerated by the shock waves produced by the supernova explosion. The shock wave from the supernova explosion expands and eventually reaches the MCs \citep{fujita2009molecular}. Subsequently, CRs escape the confinement region and begin to seep into the MCs when the escaping boundary contacts the surface of the MCs. Consequently, the hadronic component comprises the $\gamma$-ray produced from the interaction between escaped protons from SNR G40.5-0.5 and cold protons inside the associated MCs  \citep{makino2019interaction, de2022exploring}. 

We use the semi-analytical formulation by \cite{kelner2006energy} to calculate the $\gamma$-ray spectra from interactions between injected CR protons and the MC materials that surround the SNR G40.5-0.5. CR protons with an exponential cutoff power law distribution are adopted, i.e., $N_p$ $\propto$ $\gamma_{p}^{-\alpha_{p}}$e$^{-\gamma_{p}/\gamma_{p,\rm cut}}$, where $N_p$ is the number of protons in a unit volume and in the energy interval $(\gamma_p, \gamma_p + d\gamma_p)$, $\gamma_{p}$ is the proton Lorentz factor, $\alpha_{p}$ is the spectral index and $\gamma_{p,\rm cut}$ is the cutoff Lorentz factor. In this work, we invoke two scenarios to be responsible for the origin of VHE gamma-rays. The first scenario (Case 1) is that gamma rays above 100 GeV are totally attributed to the hadronic origin, and the second (Case 2) is that only gamma rays above 100 TeV are contributed by the hadronic process as a possible spectral hardening at higher energies may originate from the hadronic process \citep{2022ApJ...928..116A} and Case 1 presents a slight spectral deviation from observations above 10 TeV (see next spectral fittings).

In addition to the contributions of hadrons discussed above, we also consider the contribution from the leptonic emission of relativistic electrons in the SNR+MC system. We have considered different leptonic radiation mechanisms, such as synchrotron, bremsstrahlung and IC \citep{blumenthal1970bremsstrahlung, baring1999radio}. To calculate the IC contribution from MCs, we have considered the contribution from the interstellar radiation field (ISRF) model \citep{popescu2017radiation} and Cosmic Microwave Background (CMB) (see the left panel of Fig.~\ref{fig1}). The spectrum of the electron distribution is assumed to be a single power law with an exponential cutoff as for protons, i.e., $N_e$ $\propto$ $\gamma_{e}^{-\alpha_{e}}$e$^{-\gamma_{e}/\gamma_{e,\rm cut}}$, where $N_e$ is the number of electrons in a unit volume and in the energy interval $(\gamma_e, \gamma_e + d\gamma_e)$, $\gamma_{e}$
is the electron Lorentz factor, $\alpha_{e}$ is the index and $\gamma_{e,\rm cut}$ is the electron cutoff Lorentz factor.

\begin{deluxetable*}{cccccccccccc}
\tablenum{1}
\tablecaption{\label{tab1} Parameters used during spectral fittings for LHAASO J1908+0621.}
\tablewidth{0pt}
\tablehead{
\colhead{Component} & \colhead{Parameter} & \colhead{Fig.2 (case 1)} & \colhead{Fig.2 (case 2)}}
\startdata
Hadronic    & Spectral index ($\alpha_p$) & 1.4 & 1.6\\
                                 & Minimum energy (E$_{p,\rm min}$) & 10 GeV
& 10 GeV\\
                                 & Cutoff energy (E$_{p,\rm cut}$) & 126 TeV 
& 631 TeV\\
                                 & Total energy (W$_p$) & 1.1 $\times$ 10$^{47}$ erg
& 5 $\times$ 10$^{46}$ erg\\
                                 & Magnetic field (B) & 30 $\mu$G
& 3 $\mu$G\\
                                 & Number density (n) & 60 cm$^{-3}$
& 60 cm$^{-3}$\\
                                \hline
Leptonic  & Spectral index ($\alpha_e$) & 2.8
& 1.6\\
                                 & Minimum energy (E$_{e,\rm min}$) & 511 MeV
& 511 MeV\\
                                 & Cutoff energy (E$_{e,\rm cut}$) & 19.4 TeV
& 20.5 TeV\\
                                 & Total energy (W$_e$) & 1.6 $\times$ 10$^{48}$ erg
& 3.5 $\times$ 10$^{47}$\\
                                 & Magnetic field (B) & 30 $\mu$G
& 3 $\mu$G\\
                                 & Number density (n) & 60 cm$^{-3}$
& 60 cm$^{-3}$\\
\enddata
\end{deluxetable*}

The multi-wavelength spectral energy distribution (SED) of the source LHAASO J1908+0621 is shown in Fig.~\ref{fig2} including two different scenarios. We implement a spectral fitting and the adopted model parameters are summarized in Table \ref{tab1}. It is worth noting that in the energy range of 1-10 TeV, we give priority to the HAWC data rather than H.E.S.S. data since HAWC is better at measuring the flux of extended sources. Imaging Atmospheric Cherenkov Telescopes (IACT, e.g., H.E.S.S.) will underestimate the flux of extended sources as they usually use blank sky regions near gamma-ray sources as background emissions. However, for a large extended source, there might be dim emission in the rim taken as background, causing a lower flux measurement. In terms of the additional component in the MeV-GeV range shown in Fig.~\ref{fig2}, a consistent emission from the bremsstrahlung process is derived for Case 1, while for Case 2, the bremsstrahlung radiations move to higher energy band and therefore a direct fitting result from \cite{li2021investigating} is adopted to avoid involving the second population of electrons or protons. For two cases, the total energies of the injected protons required during spectral fittings are 1.1 $\times$ 10$^{47}$ or 5 $\times$ 10$^{46}$ erg, respectively (see Table \ref{tab1}). Both values are lower than the usual $1-10\%$ of the kinetic energy released in SNRs (typically, E$_{\rm SN}$ = 10$^{51}$ erg) \citep{aharonian2004high}. This could be attributed to the SNR age or the strength of the magnetic field in the SNR+MCs system. The adopted magnetic field strength and the MC number density are the same as in \cite{albert2022hawc}.

\subsection{LHAASO J2018+3651}

MGRO J2019+37, the MILAGRO counterpart of LHAASO J2018+3651, is one of the brightest sources in the sky at TeV energies. 
%MGRO J2019+37 detected by Milagro within the Cygnus region is towards the Cyg OB1 association. 
This source is suspected to be associated with the GeV pulsar J2021+3651 \citep{abdo2009milagro}. The estimated distance of PSR J2021+3651 ranges from 2 to 12 kpc \citep{hou2014multi}. Here we adopt a distance of 1.8 kpc as the same as in \cite{fang2020investigating} and the radius of its PWN is 24.6 pc. 
The PWN G75.2+0.1 of the pulsar PSR J2021+3651 is treated as a source of radiations of LHAASO J2018+3651.

In the PWN scenario, the leptonic origin for the multi-wavelength emissions of LHAASO J2018+3651 can be naturally expected. For the leptonic processes including synchrotron radiations and IC scatterings, the same electron distribution as used for LHAASO J1908+0621 is adopted, i.e., a single power law with an exponential cutoff. For the IC scatterings, as shown in the middle panel of Fig.~\ref{fig1}, the contributions from the ISRF model and CMB are taken into account as well.

\begin{figure*}[ht!]
\gridline{\fig{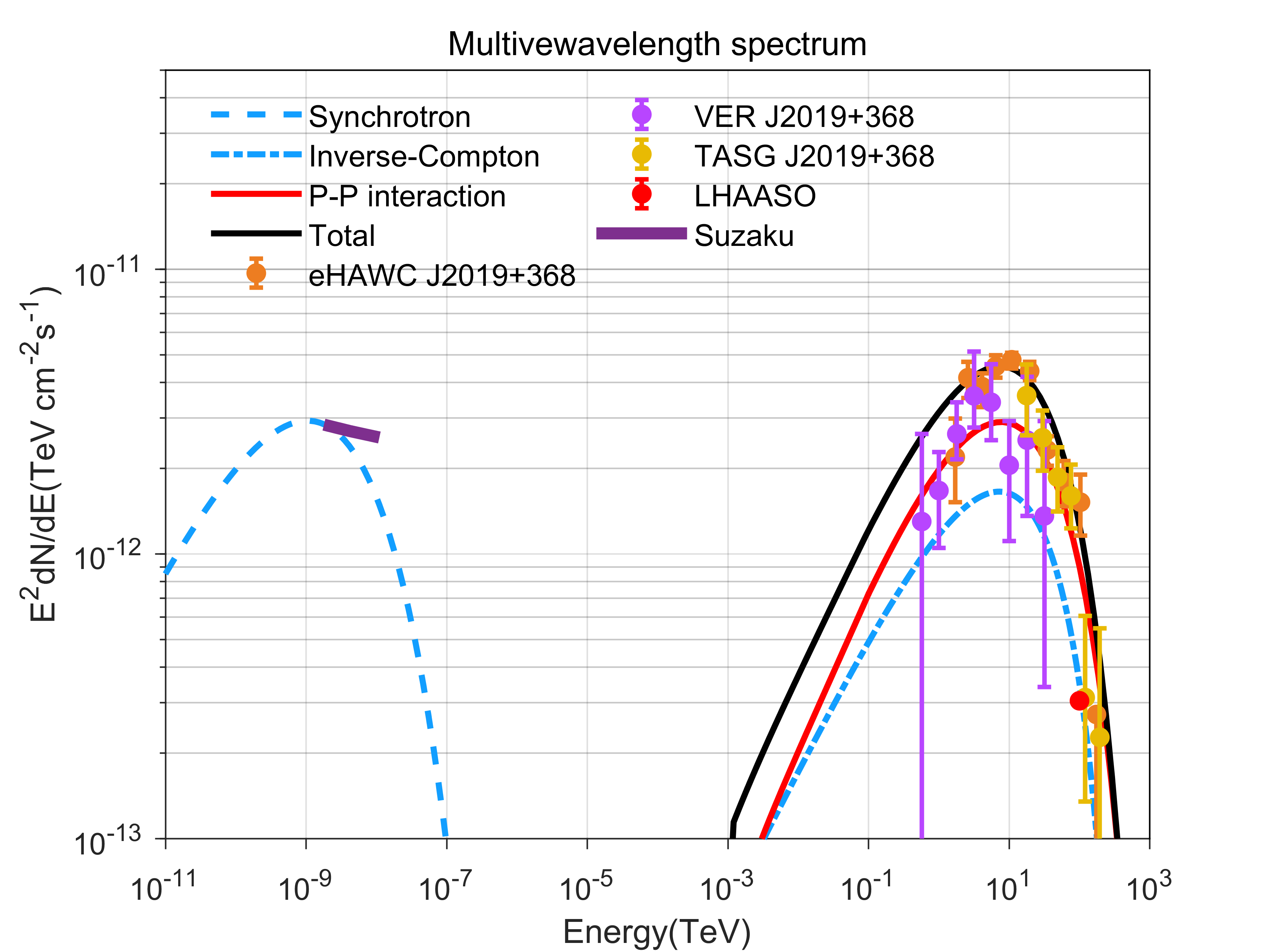}{0.49\textwidth}{(a)}
          \fig{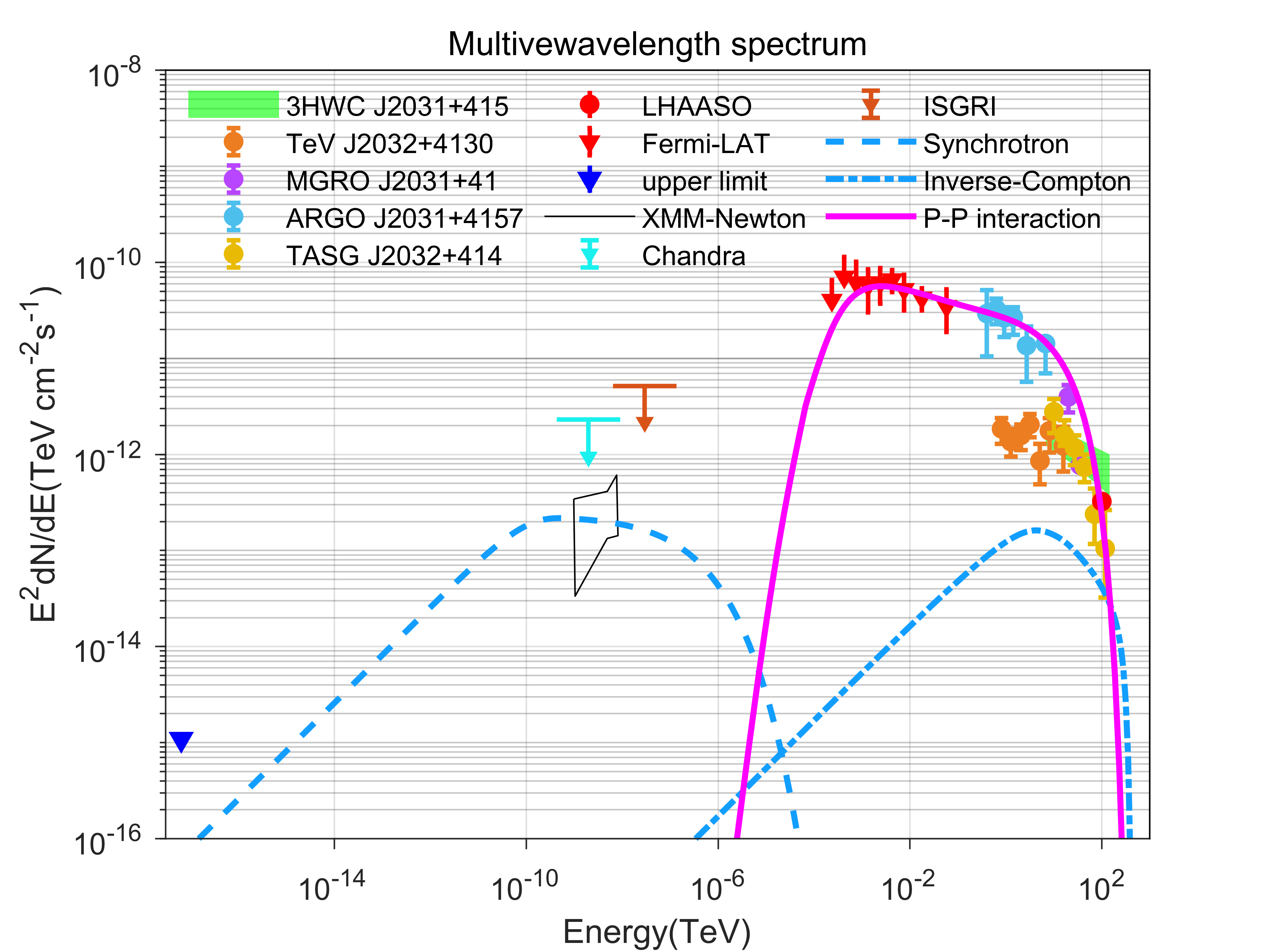}{0.49\textwidth}{(b)}
          }
          \caption{(a) Multi-wavelength SED and fitting for LHAASO J2018+3651. The synchrotron (light blue dashed) and IC (light blue dot-dashed) components are shown. The red solid line corresponds to the hadronic component from PWN G75.2+0.1 shown. The sum of IC and hadronic components is shown with a black solid line. VHE gamma-ray data points are obtained from eHAWC J2019+368 \citep{abeysekara2020multiple}, VER J2019+368 \citep{abeysekara2018very}, TASG J2019+368 \citep{amenomori2021gamma}, LHAASO \citep{cao2021ultrahigh}. The X-ray observation is obtained by \textit{Suzaku} \citep{mizuno2017x}. (b) Multi-wavelength SED and fitting for LHAASO J2032+4102. The synchrotron (light blue dashed) and IC (light blue dot-dashed) components are shown. Also, the pink solid line corresponds to the hadronic component. Gamma-ray data points are obtained from 3HWC J2031-415 \citep{albert20203hwc}, 
          TeV J2032+4130 \citep{aliu2014observations}, 
          MGRO J2031+41 \citep{abdo2007tev}, 
          ARGO J2031+4157 \citep{bartoli2014identification},
          TASG J2032+414 \citep{amenomori2021gamma}, LHAASO \citep{cao2021ultrahigh}, Fermi-LAT \citep{ackermann2011cocoon}.  The radio flux of a possible nonthermal extended radio source at $\lambda = 20$ cm \citep{paredes2006population} is assumed to be an upper limit of the actual radio emission. The upper limits between 0.5–5 keV (CX, Chandra) and 20–40 keV (ISGRI, INTEGRAL) are taken from \citep{butt2006deeper}. The model parameters of the two panels are summarized in Table~\ref{tab2}.
\label{fig3}}
\end{figure*}

In fact, apart from gamma rays produced by leptons, there may also be hadronic components involved. A portion of the pulsar's spin-down power can be converted into a stream of nuclei and nuclei can be accelerated in pulsar magnetospheres \citep{hoshino1992relativistic, arons1994relativistic, gallant1994structure}. Accelerated nuclei can undergo photodisintegration when they collide with low-energy photons generated in the nonthermal radiation fields of the pulsar's outer magnetosphere. This process results in the release of energetic neutrons that subsequently decay either within or outside the Nebula. When the protons resulting from neutron decay collide with the matter within the nebula, they generate gamma-rays and neutrinos \citep{bednarek1997gamma, liu2021pev}. Therefore, in addition to the IC process, the hadronic components can contribute $\gamma$-rays in the PWN scenario. An exponential cutoff power law distribution of protons is employed as the same as used for LHAASO J1908+0621. The multi-wavelength SED of LHAASO J2018+3651 and the spectral modeling are shown in Fig.~\ref{fig3}(a) and the adopted parameters are summarized in Table~\ref{tab2}.

Several studies have associated LHAASO J2018+3651, the counterpart of HAWC J2019+368 or MGRO J2019+37, with the PWN G75.2+0.1 powered by PSR J2021+3651~\citep{albert2021spectrum, fang2020investigating}. They explain the multiband non-thermal emission via synchrotron radiation and inverse Compton scattering in the leptonic scenario (also see \cite{woo2023hard}). \cite{hou2014multi} respectively employed the separated leptonic model or hadronic model responsible for the VHE gamma rays. Here, we adopt a lepto-hadronic model to interpret multiband emission. The ambient proton density of the PWN is still unknown \citep{beacom2007dissecting} and we adopt 1 cm$^3$, which is the same as the average density of the interstellar medium. The magnetic field is consistent with \cite{albert2021spectrum}.

\subsection{LHAASO J2032+4102}

The extended TeV gamma-ray source ARGO J2031+4157 (or MGRO J2031+41), the counterpart of LHAASO J2032+4102, is positionally coincident with the Cygnus Cocoon. No significant changes in morphology or spectrum have been observed for this extensive region \citep{ackermann2011cocoon}. However, the energy spectrum from 1 GeV to 10 TeV suggests that the Cygnus Cocoon might either be an unidentified SNR or that the particle acceleration within a superbubble is similar to that within an SNR \citep{bartoli2014identification}.

Although PSR J2032+4127 may contribute to the multiband emission of the source LHAASO J2032+4102, we choose an unknown SNR+MC system as the potential single source to power multiband observations from the LHAASO J2032+4102 region since the observed extended X-ray emission may be the counterpart of the TeV emission~\citep{bartoli2014identification}. The magnetic field $B \approx 3\,\rm \mu G$ is involved in order to be consistent with previous works \citep{horns2007xmm}. For the IC scatterings, the seed photons from the ISRF model and CMB are considered (see right panel of Fig.~\ref{fig1}). Additionally, we attribute the observed $\gamma$-rays to the decay of $\pi_0$ mesons generated through inelastic collisions between accelerated protons and target gas in an unidentified SNR+MC system as for LHAASO 1908+0621. Both the electron and proton distributions are assumed to be a power-law function with an exponential cutoff as above. The multiwavelength SED and spectral modelings of LHAASO J2032+4102 are shown in Figure~\ref{fig3}(b) and the adopted parameters are concluded in Table \ref{tab2}. The MC number density is adopted as a value of 30 cm$^{-3}$. The total energy ($\simeq W_p$) is 3 $\times$ 10$^{50}$ erg, which can be reasonably provided by one supernova, which typically releases $\sim10^{51}$ erg, and about $10\%$ of which can be transferred to the accelerated particles.

\begin{deluxetable*}{cccccccccccc}
\tablenum{2}
\tablecaption{\label{tab2} Parameters used during spectral fittings for LHAASO J2018+3651 and LHAASO J2032+4102.}
\tablewidth{0pt}
\tablehead{
\colhead{Component} & \colhead{Parameter} & \colhead{Fig.3(a) (LHAASO J2018+3651)} & \colhead{Fig.3(b) (LHAASO J2032+4102)}}
\startdata
 Hadronic    & Spectral index ($\alpha_p$) & 1.5
& 2.2\\
                               & Minimum energy (E$_{p,min}$) & 10 GeV
& 10 GeV\\
                               & Cutoff energy (E$_{p,\rm cut}$) & 316 TeV 
& 126 TeV\\
                                 & Total energy (W$_p$) & 5.5 $\times$ 10$^{48}$ erg
& 3 $\times$ 10$^{50}$ erg\\
                                 & Magnetic field (B) & 3.5 $\mu$G
& 3 $\mu$G\\
                                 & Number density (n) & 1 cm$^{-3}$
& 30 cm$^{-3}$\\
                                \hline
 Leptonic  & Spectral index ($\alpha_e$) & 2 & 2.2\\
                                 & Minimum energy (E$_{e,min}$) & 0.511 MeV
& 0.511 MeV\\
                                 & Cutoff energy (E$_{e,cut}$) & 81.6 TeV
& 205 TeV\\
                                 & Total energy (W$_e$) & 7.7 $\times$ 10$^{46}$ erg
& 2.7 $\times$ 10$^{46}$\\
                                 & Magnetic field (B) & 3.5 $\mu$G
& 3 $\mu$G\\
                                 & Number density (n) & 1 cm$^{-3}$
& 30 cm$^{-3}$\\
\enddata

\end{deluxetable*}

\begin{figure*}[ht!]
  \centering
  \includegraphics[width=0.4\textwidth]{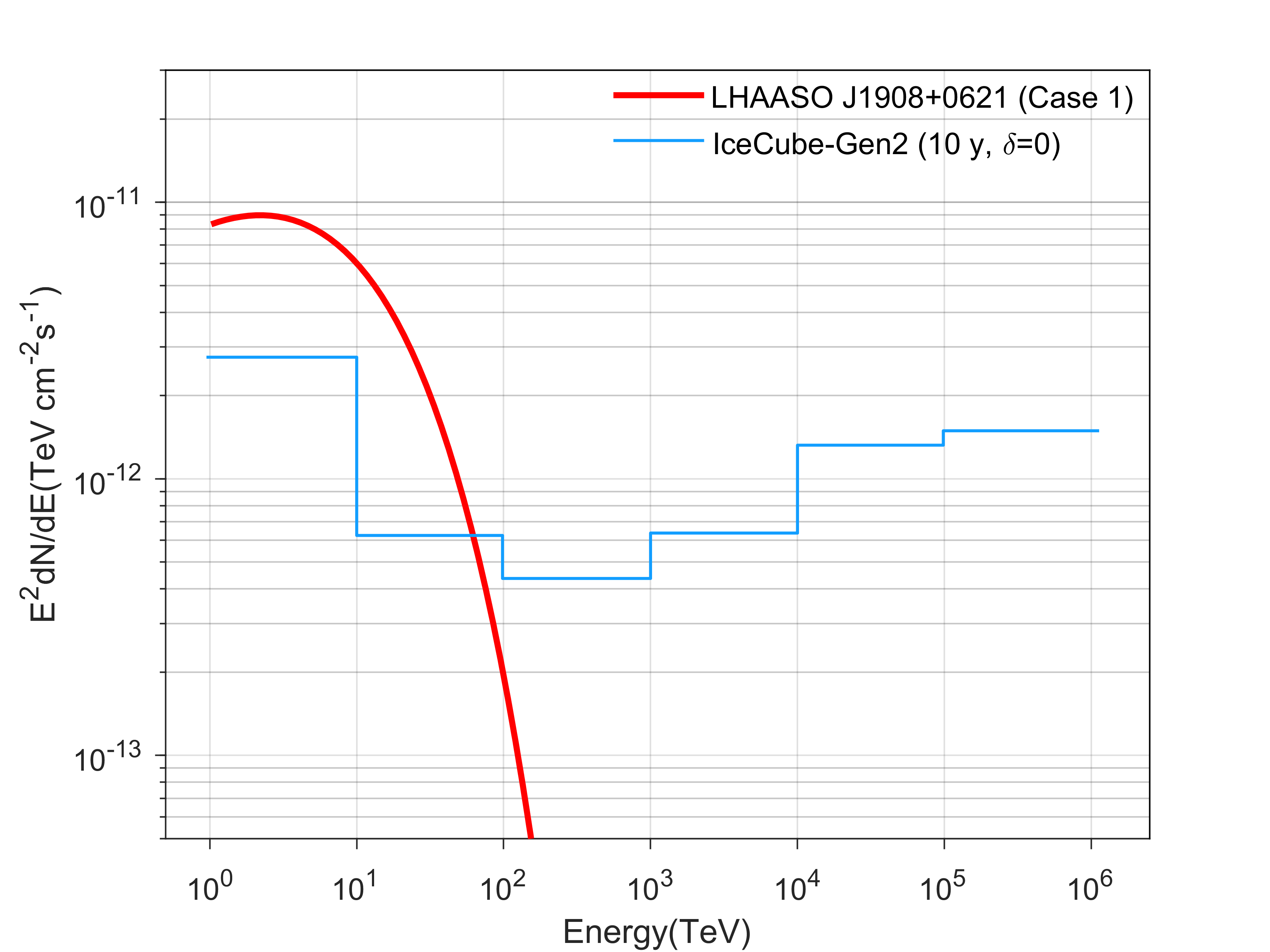}
   \hspace{0.5cm}
  \includegraphics[width=0.4\textwidth]{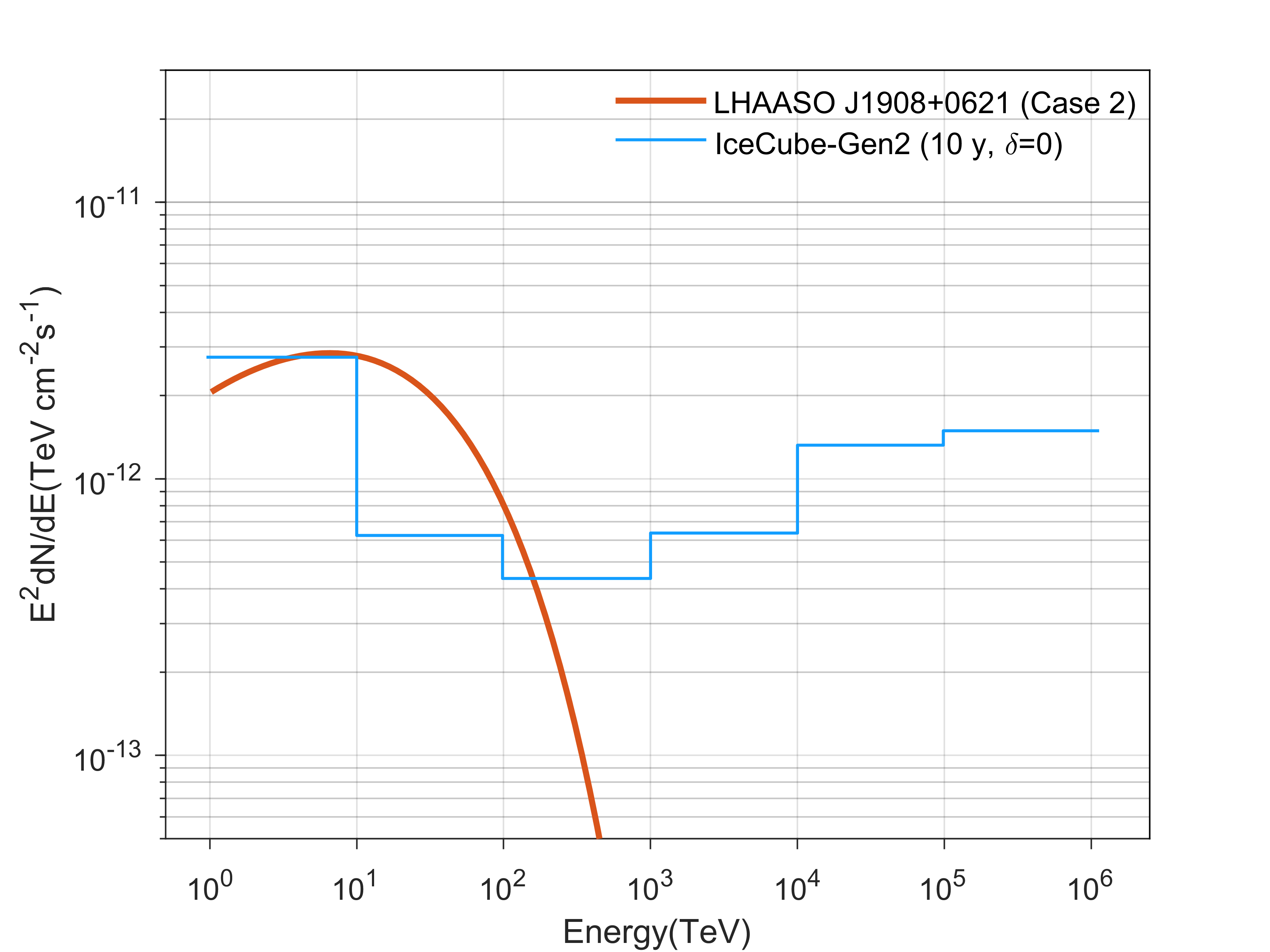}
  
  \vspace{0.3cm}
  
  \includegraphics[width=0.4\textwidth]{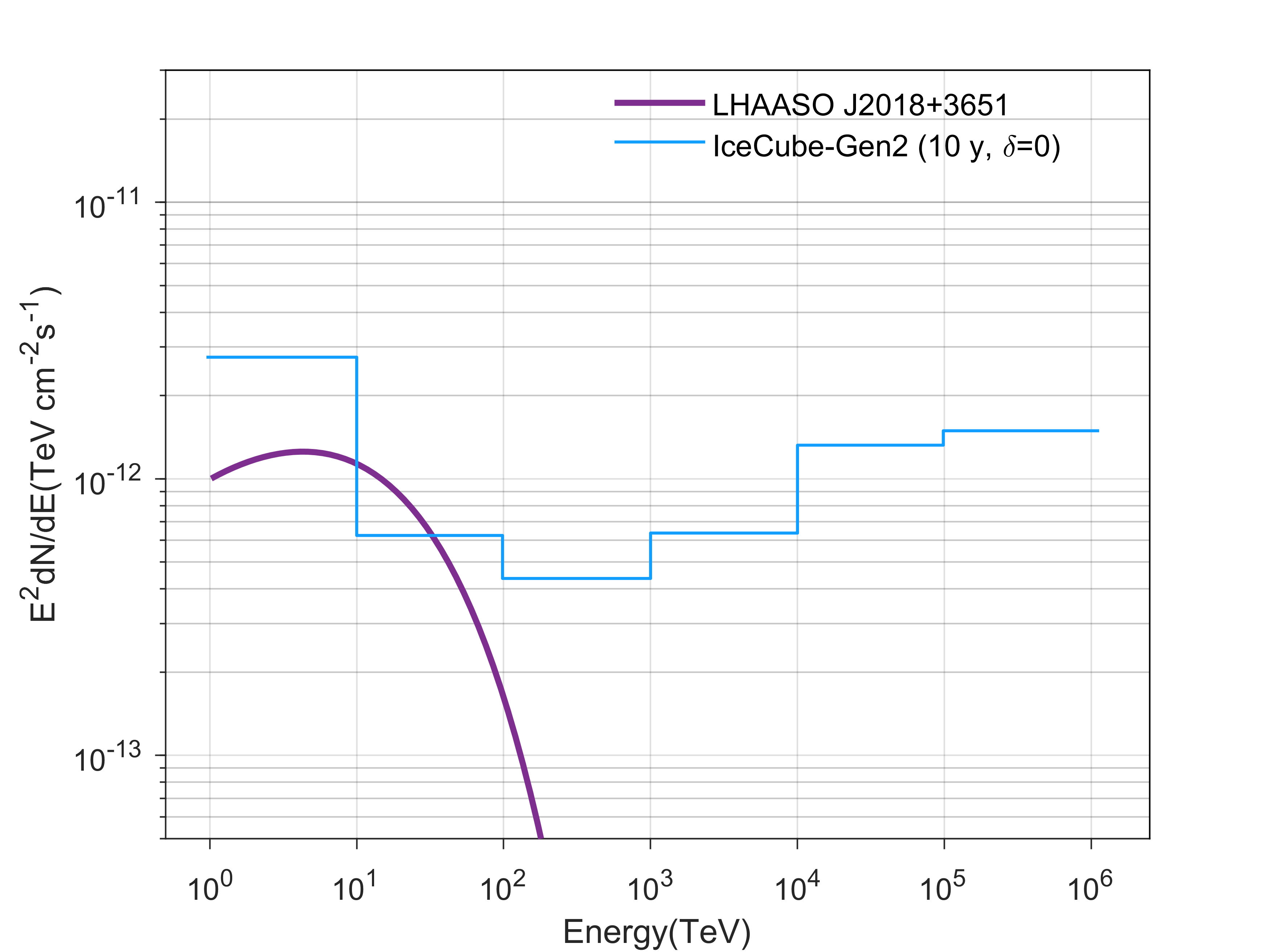}
  \hspace{0.5cm}
  \includegraphics[width=0.4\textwidth]{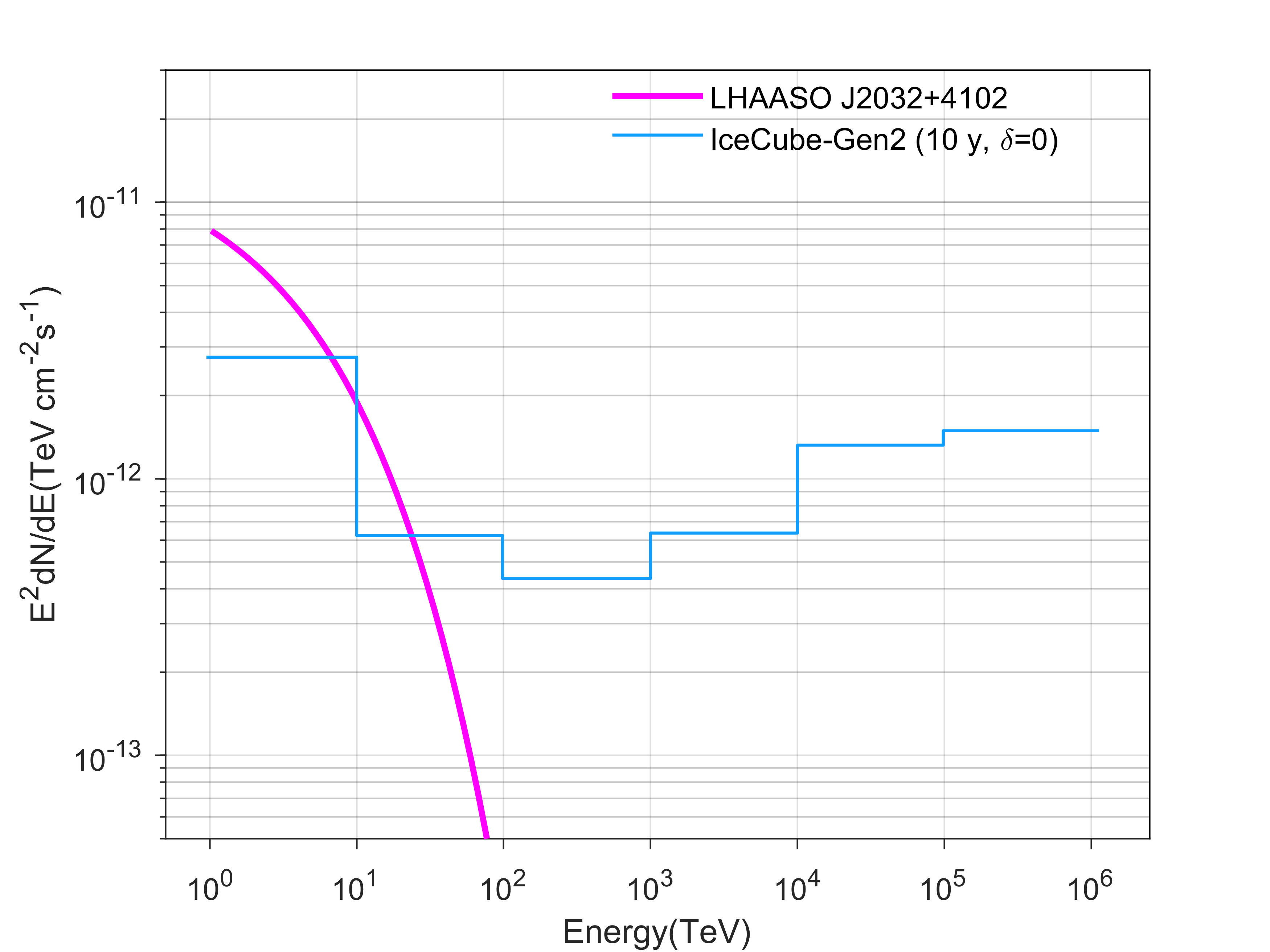}
  
  \caption{The estimated total muon neutrino flux reaching the Earth of LHAASO J1908+0621, J2018+3651, and J2032+4102, respectively. The red, brown, purple, and pink lines represent the total muon neutrino fluxes above $1\,\rm TeV$ produced by $pp$ interactions. The light blue solid line indicates the sensitivity of IceCube-Gen2 for a point source at the celestial equator with an average significance of 5$\sigma$ after 10 years of observations \citep{aartsen2019neutrino}.}
  \label{fig4}
\end{figure*}

% \begin{figure*}[ht!]
%     \includegraphics[width=\textwidth]{p-value.jpg}
%     \caption{The statistical significance (p-value) as a function of observation time for three sources. The blue and orange lines represent the p-values obtained using the IceCube and IceCube-Gen2 respectively. The starting point of the blue line is the ten-year (2008-2018) neutrino source search by IceCube.}
%     \label{fig5}
% \end{figure*}

\begin{figure*}[ht!]
  \centering
  \includegraphics[width=1\textwidth]{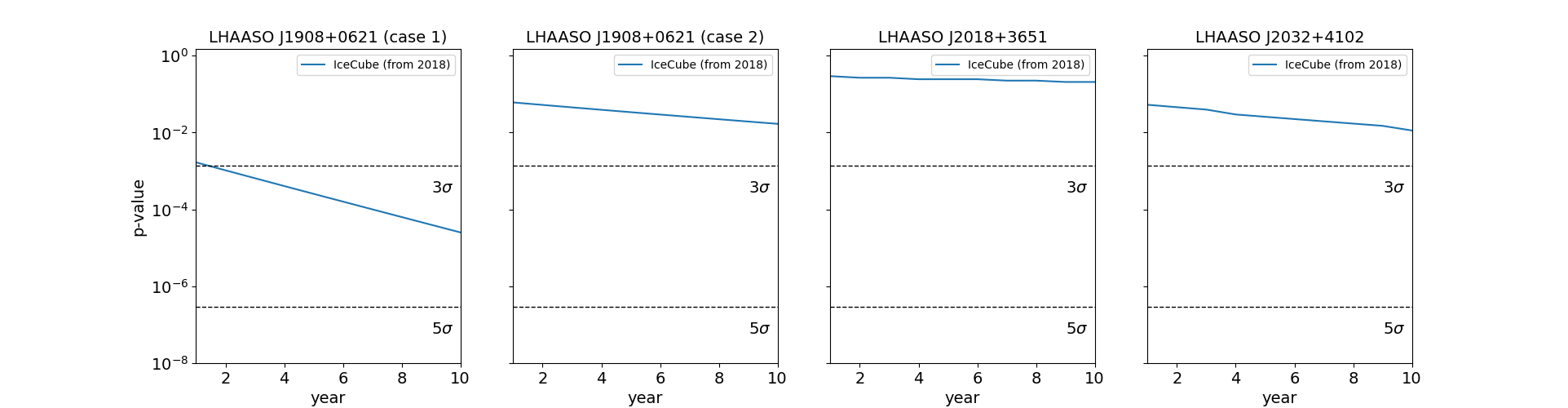}
  
  \vspace{0.1cm}
  
  \includegraphics[width=1\textwidth]{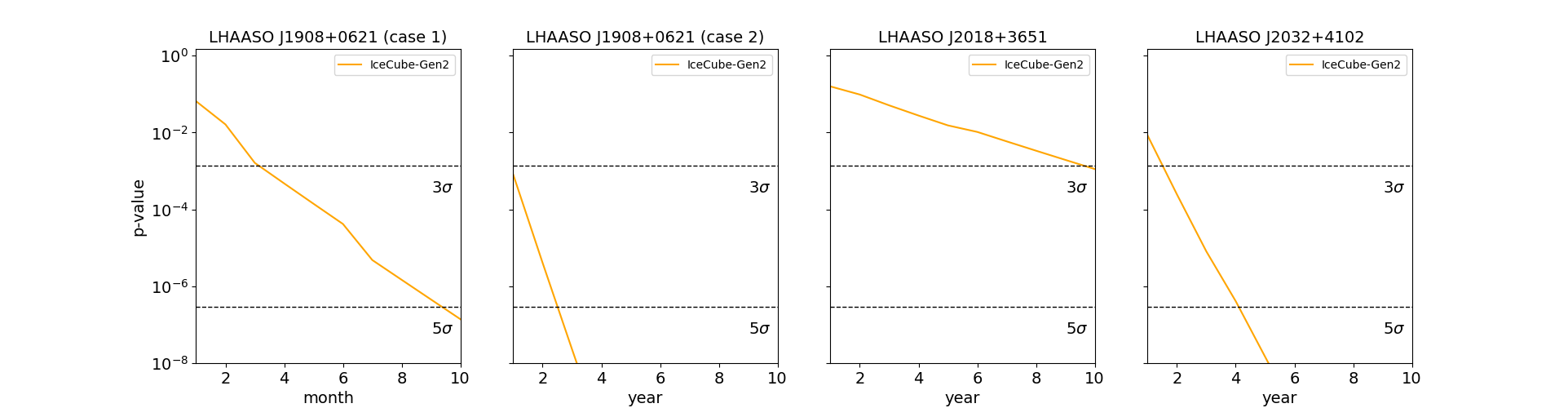}
  
  \caption{The statistical significance (p-value) as a function of observation time for three sources. 
 The blue lines show the expected p-values obtained using IceCube in addition to the ten years from 2008 to 2018. The orange lines show the expected p-values obtained using IceCube-Gen2.}
  \label{fig5}
\end{figure*}

\section{Neutrino flux}\label{sec:neu}

Neutrinos are produced alongside $\gamma$-rays in hadronic $pp$ interactions. Therefore, if there are $\gamma$-ray sources powered by hadronic interactions, it is also expected that neutrinos will be emitted from the same source region.
%MGRO J1908+06, MGRO J2019+37 and MGRO J2031+41, which are the MILAGRO counterparts of LHAASO J1908+0621, LHAASO J2018+3651, and LHAASO J2032+4102, may be neutrino sources due to their extended nature and hard TeV $\gamma$-ray spectrum \citep{gonzalez2009identifying, halzen2017prospects}. IceCube neutrino telescope has conducted searches for point-like source emissions in the vicinity of these sources, which further supports this possibility.
To calculate the flux of muon neutrinos reaching Earth from the three sources, we use the semi-analytical formulation developed in \cite{kelner2006energy}. The derived muon neutrino flux is shown in Fig.~\ref{fig4} after considering neutrino oscillations. Neutrino oscillations induce an equal flux among different neutrino flavors when they reach Earth, namely, the (anti-)muon neutrino flux reaching Earth is one-third of all-flavor neutrino flux, i.e., $(\Phi_{\nu_{\mu}}+\Phi_{\overline{\nu}_{\mu}}+\Phi_{\nu_e}+\Phi_{\overline{\nu}_e})/3$ with produced neutrino fluxes $(\Phi_{\nu_{\mu}},\Phi_{\overline{\nu}_{\mu}},\Phi_{\nu_e},\Phi_{\overline{\nu}_e})$ by pp interactions at the source.

As shown in Fig.~\ref{fig4}, the model predicts a neutrino flux that exceeds the sensitivity limit of next-generation IceCube-Gen2. This indicates that if the total observed TeV/PeV gamma-ray emission from these LHAASO sources originates from the hadronic processes, the accompanying neutrino flux can be identified by IceCube-Gen2. As no accompanying neutrinos will be expected if VHE gamma rays are produced by leptonic processes, the detection or non-detection by IceCube-Gen2 in the future can provide further insights and help to determine the origin of VHE gamma rays from these LHAASO sources.

\section{FUTURE PROSPECTS \label{sec:future}}

In this section, we evaluate the detection of neutrinos from three sources by IceCube and IceCube-Gen2. The number of signal events from the point-like source at the declination $\delta_s$ is expressed as
\begin{equation}
    n_{s}= t \int \mathrm{d}E_{\nu} \frac{\mathrm{d}N_{\nu}(E_{\nu})}{\mathrm{d}E_{\nu}} A_{\rm eff}(E_{\nu}, \delta_s),
\end{equation}
where $A_{\rm eff}$ is the effective area for through-going track events. The background events from the three sources are mainly induced by atmospheric neutrinos. Thus, the number of background events is expressed as
\begin{equation}
    n_{b}= t \int \mathrm{d}\Omega \int \mathrm{d}E_{\nu} (I_{\nu,\rm atm} + I_{\nu,\rm astro}) A_{\rm eff}(E_{\nu}, \theta_{z}),
\end{equation}
where $I_{\nu,\rm atm}$ is the atmospheric muon neutrino flux calculated by MCEq \citep{2015EPJWC..9908001F}, and $I_{\nu,\rm astro}$ is the diffuse astrophysical muon neutrino flux \citep{2022ApJ...928...50A}. As for IceCube, the effective area and the smearing matrix are released with the ten-year (2008-2018) muon track data \citep{icpstrack2018}. The smearing matrix gives the reconstructed energy distribution for neutrinos with different energies and incident directions. As for IceCube-Gen2, the effective area is assumed to be 7.5 times that of IceCube \citep{2022icrc.confE1185S}, and the smearing matrix is assumed to be the same as that of IceCube. Low-energy cut is applied to both IceCube and IceCube-Gen2 observations, aiming for the neutrino-induced muons with reconstructed energy $E_{\rm rec}<50\,{\rm TeV}$ \citep{IceCube-Gen2:2021tmd}.

We estimate the statistical significance of observation with a p-value analytically expressed as \citep{CMS-NOTE-2011-005, halzen2017prospects}
\begin{flalign}
p_{\text {value}}&=\frac{1}{2}\left[1-\operatorname{erf}\left(\sqrt{q_{0}^{\rm obs} / 2}\right)\right],
\end{flalign}
where
\begin{flalign}
q_{0}^{\rm obs}&=2\left[Y_{b}-N_{D}+N_{D} \ln \left(\frac{N_{D}}{Y_{b}}\right)\right],
\end{flalign}
$Y_{b}$ is the expected number of background events, and $N_{D}$ is the median of Poisson-distributed events containing both signal and background. The event numbers are counted within the solid angle $\Omega=1.6\sigma$, where $\sigma$ is the angular resolution of the detector. This solid angle contains 72\% of the signal from a point-like source \citep{1993NIMPA.328..570A}. The angular resolution $\sigma$ of IceCube is assumed to be the median angular uncertainty of the muon-track events ($E_{\rm rec}<50\,{\rm TeV}$) observed within the declination band $\delta_s\pm1^{\circ}$.
\cite{IceCube-Gen2:2021tmd} reports the angular resolutions for low-energy samples simulated by IceCube-Gen2 (see Table 2 therein), but only three zenith bins are offered. Thus, we extrapolate the angular resolution to the source declination $\delta_s$ as $\sigma(\delta_s)=\sigma_{0}(\delta_s)+0.05^{\circ}$,  where $\sigma_0$ represents the angular resolution of IceCube-Gen2 for 10 TeV muons (see  Figure 24 in \cite{IceCube-Gen2:2020qha}). 

The results for the statistical significance are reported in Fig.~\ref{fig5}. With the observation by IceCube, LHAASO J1908+0621 is expected to reach $3\sigma$ in $\sim$2020 ($\sim$2 years after 2018) for Case 1. In the latest observation by IceCube (2008-2020)~\citep{2023ApJ...945L...8A}, the pre-trial p-value for LHAASO J1908+0621 is only 0.046 under the assumption of a power-law spectrum. Therefore, our results tend to disfavor Case 1 for LHAASO J1908+0621, which corresponds to that gamma rays above $100\,\rm GeV$ are of purely hadronic origin.
LHAASO J1908+0621 will be detected at 5$\sigma$ level in less than 10 months for Case 1, while for Case 2 will be detected at 5$\sigma$ level in less than 3 years with IceCube-Gen2 detector, considering the relevant parameters as reported in Table \ref{tab1}. Therefore, future observations by the next generation neutrino telescope can further help distinguish the origin of VHE gamma-rays from LHAASO J1908+0621. The source LHAASO J2018+3651 and LHAASO J2032+4102 could be detected at 3$\sigma$ level in $\sim$10 years and 5$\sigma$ level in $\sim$4 years by IceCube-Gen2, respectively, considering the relevant parameters as reported in Table \ref{tab2}. However, no significant statistical significance will be expected by IceCube for LHAASO J2018+3651 and LHAASO J2032+4102 in the next 10 years.

\section{discussion and summary}\label{sec:discussion and summary}

Galactic high-energy neutrinos have been long expected. Galactic sources detected by LHAASO are thought of as the potential PeVatrons since gamma rays with energies larger than 100 TeV have been found from them. However, it is still under debate whether VHE gamma rays are produced by the leptonic process or the hadronic process. Neutrino observations can be an important probe to distinguish the origin of VHE gamma rays since the hadronic process ($pp$ collisions or photomeson production process) will produce the accompanying high-energy neutrinos. In this paper, we investigate multiband spectra of three LHAASO sources, i.e., LHAASO J1908+0621, LHAASO J2018+3651, and LHAASO J2032+4102, which are the most promising galactic neutrino sources. We propose reasonable lepto-hadronic scenarios to implement the multiband spectral modeling. 

Assuming gamma rays are entirely hadronic, we calculate the most optimistic flux of muon neutrinos generated from the hadronic process. Furthermore, we estimate the statistical significance (p-value) as a function of time for three sources using both the IceCube neutrino observatory and the proposed second-generation IceCube-Gen2. Our results tend to disfavor that all gamma rays above $100\,\rm GeV$ (Case 1) from LHAASO J1908+0621 are of purely hadronic origin based on current IceCube observations. However, the purely hadronic origin of gamma rays above $100\,\rm TeV$ (Case 2) from LHAASO J1908+0621 is still possible. No significant statistical significance will be expected by IceCube for LHAASO J2018+3651 and LHAASO J2032+4102 in the next 10 years. Besides, for IceCube-Gen2, LHAASO J1908+0621 can be detected at a 5$\sigma$ level within 10 months for Case 1, and within 3 years for Case 2. Similarly, high-energy neutrinos from LHAASO J2018+3651 and LHAASO J2032+4102 can be respectively detected by IceCube-Gen2 at a 3$\sigma$ level in $\sim$10 years and a 5$\sigma$ level in $\sim$4 years if the VHE gamma rays are entirely hadronic. Future observations by IceCube-Gen2 or other more advanced next-generation neutrino telescopes at the positions of three sources will be important to untangle the exact nature of these enigmatic sources.

\section*{Acknowlegdments}

We thank the anonymous referee for the helpful comments. We acknowledge support from the National Natural Science Foundation of China under grant No.12003007 and the Fundamental Research Funds for the Central Universities (No. 2020kfyXJJS039)

\bibliography{Reference}{}
\bibliographystyle{aasjournal}

\end{document}